\documentclass[preprint,aps]{revtex4}

\usepackage{graphics}

\begin{document}

\title{Asymptotic Quasinormal Frequencies of Different Spin Fields in Spherically Symmetric Black Holes}

\author{H. T. Cho}
  \email{htcho@mail.tku.edu.tw}
\affiliation{Department of Physics, Tamkang University, Tamsui,
Taipei, Taiwan, Republic of China}

\date{\today}

\begin{abstract}
We consider the asymptotic quasinormal frequencies of various spin
fields in Schwarzschild and Reissner-Nordstr\"om black holes. In
the Schwarzschild case, the real part of the asymptotic frequency
is ln3 for the spin 0 and the spin 2 fields, while for the spin
1/2, the spin 1, and the spin 3/2 fields it is zero. For the
non-extreme charged black holes, the spin 3/2 Rarita-Schwinger
field has the same asymptotic frequency as that of the integral
spin fields. However, the asymptotic frequency of the Dirac field
is different, and its real part is zero. For the extremal case,
which is relevant to the supersymmetric consideration, all the
spin fields have the same asymptotic frequency, the real part of
which is zero. For the imaginary parts of the asymptotic
frequencies, it is interesting to see that it has a universal
spacing of $1/4M$ for all the spin fields in the single-horizon
cases of the Schwarzschild and the extreme Reissner-Nordstr\"om
black holes. The implications of these results to the universality
of the asymptotic quasinormal frequencies are discussed.
\end{abstract}

\pacs{04.70.-s, 11.15.Kc, 11.30.Pb}

\maketitle

\section{Introduction}

Hod \cite{hod1} was the first to conjecture that the highly damped
limit of the black hole quasinormal frequency was related to the
fundamental area unit in the quantum theory of gravity. At that
time, this limit was known only numerically
\cite{nollert,andersson},
\begin{equation}
\omega_{n}=0.0437123\left(\frac{1}{M}\right)
-\frac{i}{4M}\left(n+\frac{1}{2}\right)+\cdots,
\end{equation}
as $n\rightarrow\infty$, where $M$ is the mass of the black hole.
He noticed that 0.0437123 is very close to ${\rm ln}3/8\pi$. Using
Bohr's correspondence principle, he was able to derive the area
spectrum of the quantum Schwarzschild black hole to be
\begin{equation}
A_{n}=4({\rm ln}3)n,\ \ \ n=1,2,3,\dots
\end{equation}
Comparing Hod's result with the expressions of the area and
entropy spectra obtained in the theory of loop quantum gravity,
Dreyer \cite{dreyer} determined the value of the Immirzi
parameter, an otherwise arbitrary constant in the theory. At the
same time, because of the presence of $\ {\rm ln}3$, he also
suggested that the gauge group should be changed from SU(2) to
SO(3). Although this connection between the asymptotic quasinormal
frequency and the Immirzi parameter has been questioned
\cite{domagala,chou}, it has nevertheless aroused a lot of
research interests in this direction.

The first analytic evaluation of the asymptotic quasinormal
frequency was carried out by Motl and Neitzke \cite{motl,neitzke}
using the monodromy method. Subsequently, with this method, the
calculation has been extended to other kinds of black holes
\cite{natario}. However, all these calculations are done with
respect to fields with integral spins. In this paper, we would
like to further consider the asymptotic quasinormal frequencies of
fields with half-integral spins like the Dirac and the
Rarita-Schwinger fields, which are lacking so far. On the other
hand, we hope our consideration will also shed light on the
question of universality of the value of $\ {\rm ln}3$ studied by
several authors \cite{tamaki,kettner,das}. It turns out that this
value is indeed obtained in most of the cases for single-horizon
black holes. We would like to see if this universality can be
applied to fields with different spins.

In the next section, we first consider the case of the
Schwarzschild black hole. To deal with different spin fields in a
unified way, we use the WKB formalism of Andersson and Howls
\cite{howls}, in addition to the monodromy method, to evaluate the
asymptotic frequencies. Here we also address the discrepancy on
the value of the imaginary part of the Dirac asymptotic frequency
in \cite{castello} and \cite{jing}. In Section III, we turn to the
case of the charged Reissner-Nordstr\"om black hole. We shall
consider both the non-extremal and the extremal black holes. Since
the horizon structures are different in these two cases, one
cannot take the extremal limit directly. A separate calculation is
thus carried out carefully in this section. In so doing, we would
also hope to resolve the discrepancies in the value of the
asymptotic quasinormal frequencies for the extremal black holes in
the literature \cite{natario,das,howls}. Moreover, this
calculation is also interesting because the spin 1, the spin 3/2,
and the spin 2 fields together in the extremal black hole
spacetime represent the simplest supersymmetric situation. We
would like to see how their asymptotic frequencies are related in
this case \cite{onozawa}. Conclusions and discussions are
presented in Section IV.

\section{Schwarzschild black hole}

For the Schwarzschild black hole, the metric can be written as,
\begin{equation}
ds^{2}=-\frac{\Delta}{r^{2}}dt^{2}+\frac{r^{2}}{\Delta}dr^{2}+r^{2}d\Omega^{2},
\label{metric}
\end{equation}
where $\Delta=r(r-2M)$ and $M$ is the mass of the black hole. The
radial parts of the wave equations for different spin fields can
all be simplified to the form of a Schr\"odinger-like equation,
\begin{equation}
\frac{d^{2}Z(r)}{dr_{*}^{2}}+(\omega^{2}-V)Z(r)=0,
\label{schrodinger}
\end{equation}
where $\omega$ is the frequency, $V$ is the effective potential,
and $r_{*}$ is the so-called tortoise coordinate with
\begin{equation}
\frac{d}{dr_{*}}=\left(\frac{\Delta}{r^{2}}\right)\frac{d}{dr}
\Rightarrow r_{*}=r+2M{\rm ln}\left(\frac{r}{2M}-1\right).
\end{equation}

For integral spins, $s=0,1,$ and $2$ \cite{chandrasekhar},
\begin{equation}
V=\frac{\Delta}{r^{2}}\left[\frac{l(l+1)}{r^{2}}+\frac{(1-s^{2})2M}{r^{3}}\right],
\end{equation}
where $l=0,1,2,\dots$ is the angular momentum number. For the
Dirac field, $s=1/2$ \cite{cho},
\begin{equation}
V=\frac{\Delta^{1/2}}{r^{4}}\left[\kappa^{2}\Delta^{1/2}-\kappa(r-3M)\right],
\end{equation}
where $\kappa=j+1/2$ and $j=l\pm 1/2$, $l=0,1,2,\dots$. For the
Rarita-Schwinger field, $s=3/2$ \cite{aichburg,torres},
\begin{equation}
V=\frac{\Delta}{r^{6}}(\lambda r^{2}+2Mr)-\frac{d}{dr_{*}}
\left[\frac{1}{F}\left(\frac{dF}{dr_{*}}-\lambda\sqrt{\lambda+1}\right)\right],
\end{equation}
where
\begin{equation}
F=\frac{1}{\Delta^{1/2}}(\lambda r^{2}+2Mr),
\end{equation}
and
\begin{equation}
\lambda=\left(j-\frac{1}{2}\right)\left(j+\frac{3}{2}\right),
\end{equation}
with $j=l+3/2$, $l=0,1,2,\dots$. We have listed the effective
potentials for various spin fields here for completeness. In the
following calculations, we need mainly the asymptotic behaviors of
these potentials as $r\rightarrow 0$, that is, near the black hole
singularity. We assume that as $r\rightarrow 0$, the asymptotic
behavior of the effective potential is
\begin{equation}
V\sim\frac{M^{2}}{r^{4}}\alpha.\label{SchV}
\end{equation}
The values of $\alpha$ for various spin fields are listed in
Table~\ref{table1}.

\begin{table}
\caption{\label{table1}Some parameters characterizing the
asymptotic behaviors of various spin fields for the Schwarzschild
black hole. $\alpha$ represents the behavior of the effective
potential near the black hole singularity as in Eq.~(\ref{SchV}).
$\gamma$ is the value of the integral in Eq.~(\ref{Schgamma}).}
\begin{ruledtabular}
\begin{tabular}{cccc}
spin&$\alpha$&$\gamma$&$-(1+2\ \!{\rm cos}2\gamma)$\\ \hline 0 &
$-4$ & 0 & $-3$ \\ 1/2 & 0 & $-\pi$/2 & 1 \\ 1 & 0 & $-\pi$/2 & 1
\\ 3/2 & 0 & $-\pi$/2 & 1 \\ 2 & 12 & $-\pi$ & $-3$ \\
\end{tabular}
\end{ruledtabular}
\end{table}

Going back to Eq.~(\ref{schrodinger}), the solutions at infinity,
$r_{*}\rightarrow\pm\infty$, are
\begin{equation}
Z(r)\sim e^{\pm i\omega r_{*}},
\end{equation}
because $V\rightarrow 0$ in this limit. The quasinormal modes
correspond to the solutions with the boundary conditions of
outgoing wave, $e^{i\omega r_{*}}$, at $r_{*}=r=\infty$, and
ingoing wave, $e^{-i\omega r_{*}}$, at the horizon,
$r_{*}=-\infty$ or $r=2M$. The corresponding spectrum of these
modes are complex and discrete.

In order to use the WKB method of Andersson and Howls to evaluate
the asymptotic quasinormal frequencies as $|{\rm
Im}\omega|\rightarrow\infty$, we define a new function
\cite{howls},
\begin{equation}
\psi(r)=\frac{\Delta^{1/2}}{r}Z(r).
\end{equation}
From Eq.~(\ref{schrodinger}), one can write the wave equation for
$\psi$ as
\begin{equation}
\frac{d^{2}\psi}{dr^{2}}+R(r)\psi=0,\label{WKBeq}
\end{equation}
with
\begin{equation}
R(r)=\frac{r^{2}}{(r-2M)^{2}}\left[\omega^{2}-V
+\frac{2M}{r^{3}}-\frac{3M^{2}}{r^{4}}\right].
\end{equation}
The WKB solutions to this equation are \cite{howls},
\begin{equation}
f_{1,2}^{(t)}(r)=\frac{1}{\sqrt{Q(r)}}e^{\pm i\int_{t}^{r}d\xi
Q(\xi)},\label{WKB}
\end{equation}
where $t$ is a reference point and
\begin{eqnarray}
Q^{2}(r)&=&R(r)-\frac{1}{4r^{2}}\nonumber\\
&=&\frac{r^{2}}{(r-2M)^{2}}\left[\omega^{2}-V-\frac{1}{4r^{2}}
+\frac{3M}{r^{3}}-\frac{4M^{2}}{r^{4}}\right].\label{SchQ}
\end{eqnarray}
Here $Q(r)$ is chosen in such a way to match the behavior of the
solutions of $\psi(r)$ near $r=0$.

\begin{figure}[!]
\includegraphics{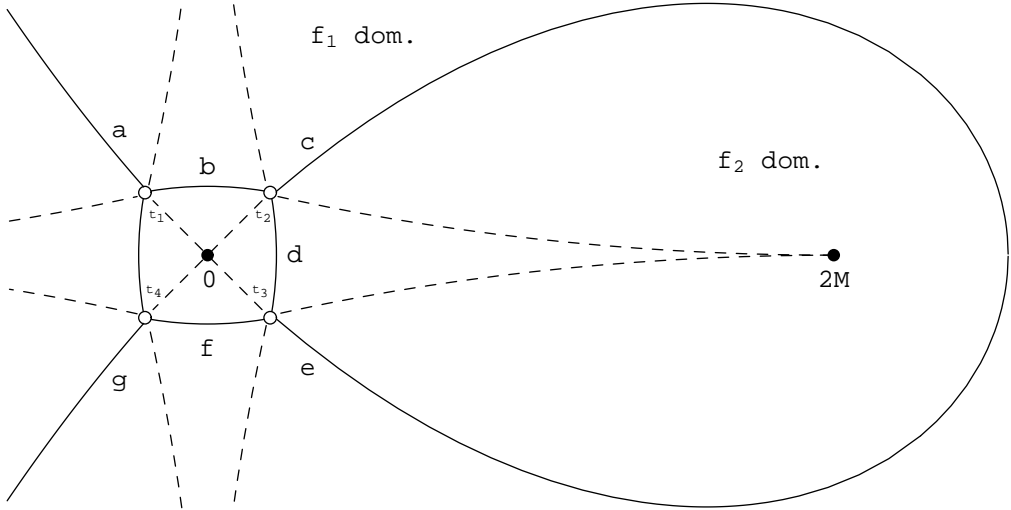}
\caption{\label{schwarzschild}Stokes structure of the
Schwarzschild black hole. Open circles are zeros of $Q(r)$ and
filled circles are poles of $Q(r)$. The poles are located at the
black hole singularity ($r=0$) and the event horizon ($r=2M$).
Solid lines are anti-Stokes lines and broken lines are Stokes
lines. The regions where $f_{1}$ or $f_{2}$ dominates are also
indicated.}
\end{figure}

The zeros and the poles of $Q(r)$ are important to the behaviors
of the WKB solutions $f_{1,2}^{(t)}(r)$. In the Schwarzschild
case, as shown in Fig.~\ref{schwarzschild}, $Q(r)$ has four zeros.
From each zero, three Stokes lines and three anti-Stokes lines
emanate. Along the anti-Stokes lines $Q(r)dr$ is purely real, so
$f_{1}^{(t)}(r)$ and $f_{2}^{(t)}(r)$ are oscillatory functions of
comparable magnitudes. Between anti-Stokes lines are regions on
the complex $r$-plane in which one of the two WKB solutions
dominates. We have also indicated this in
Fig.~\ref{schwarzschild}. On the Stokes lines $Q(r)dr$ is purely
imaginary. There are also two poles at $r=0$ and at $r=2M$. The
solution to the wave equation in the WKB approximation is
represented by an appropriate combination of $f_{1}^{(t)}(r)$ and
$f_{2}^{(t)}(r)$. The behavior of this solution changes as one
crosses the Stokes lines. This is the so-called Stokes phenomenon.
By incorporating these changes, one can derive the asymptotic
behavior of the solution on the whole complex plane.

To start the calculation, we consider the boundary condition of
the quasinormal mode at spatial infinity. Assuming that Re\
$\omega>0$, one can analytically continue this boundary condition
to the anti-Stokes line labelled $a$ in Fig.~\ref{schwarzschild}.
With the definition of the WKB solutions in Eq.~(\ref{WKB}), the
boundary condition at $a$ becomes
\begin{equation}
\psi_{a}=f_{1}^{(t_{1})},
\end{equation}
where we have indicated explicitly from which zero the anti-Stokes
line emanates. Going to $b$ in the clockwise direction, we cross a
Stokes line. Since this Stokes line locates in a region where
$f_{1}$ dominates, the $f_{1}$ part of $\psi_{a}$ will not change
but there will be an additional $f_{2}$ part with the coefficient
of $f_{1}$ in $\psi_{a}$ (which is 1 here) multiplying $-i$ for
crossing the line in the clockwise direction. (If we had crossed
the Stokes line in the counterclockwise direction, we would have
to multiply by $i$ instead.) Hence, at $b$,
\begin{equation}
\psi_{b}=\psi_{a}-if_{2}^{(t_{1})}=f_{1}^{(t_{1})}-if_{2}^{(t_{1})}.
\end{equation}
Next, we have to change the reference point from $t_{1}$ to
$t_{2}$.
\begin{equation}
f_{1,2}^{(t_{1})}=e^{\pm i\gamma_{12}}f_{1,2}^{(t_{2})},
\end{equation}
where
\begin{equation}
\gamma_{12}=\int_{t_{1}}^{t_{2}}d\xi\ Q(\xi)\equiv
\gamma.\label{Schgamma}
\end{equation}
Now near the zeros, $r$ is small because Im\ $\omega\rightarrow
-\infty$,
\begin{equation}
Q^{2}(r)\approx\frac{r^{2}}{4M^{2}}\left[\omega^{2}
-\frac{M^{2}}{r^{4}}(\alpha+4)\right],
\end{equation}
from Eq.~(\ref{SchQ}). Taking $y=\xi^{2}\omega/M\sqrt{4+\alpha}$,
we have
\begin{eqnarray}
\gamma&=&\int_{t_{1}}^{t_{2}}d\xi\left(\frac{\omega\xi}{2M}\right)
\left[1-\frac{M^{2}(\alpha+4)}{\omega^{2}\xi^{4}}\right]^{1/2}
=\frac{\sqrt{4+\alpha}}{4}\int_{-1}^{1}dy
\left(1-\frac{1}{y^{2}}\right)^{1/2}\nonumber\\
&=&-\frac{\pi}{2}\sqrt{1+\frac{\alpha}{4}}.
\end{eqnarray}
One can also show that
$\gamma_{12}=-\gamma_{23}=\gamma_{34}=\gamma$. The value of
$\gamma$ here is crucial in the derivation of the asymptotic
quasinormal frequency. They are also listed in Table~\ref{table1}.

After changing the reference point to $t_{2}$,
\begin{equation}
\psi_{b}=e^{i\gamma}f_{1}^{(t_{2})}-ie^{-i\gamma}f_{2}^{(t_{2})}.
\end{equation}
Going to $c$, we cross another anti-Stokes line, so
\begin{eqnarray}
\psi_{c}&=&\psi_{b}-i(e^{i\gamma})f_{2}^{(t_{2})}\nonumber\\
&=&e^{i\gamma}f_{1}^{(t_{2})}-i(e^{i\gamma}+e^{-i\gamma})f_{2}^{(t_{2})}.
\end{eqnarray}
Going to $d$, we cross yet another anti-Stokes line. However, we
are in a region where $f_{2}$ dominates. Hence,
\begin{eqnarray}
\psi_{d}&=&\psi_{c}-i(-ie^{i\gamma}-ie^{-i\gamma})f_{1}^{(t_{2})}\nonumber\\
&=&-e^{-i\gamma}f_{1}^{(t_{2})}-i(e^{i\gamma}+e^{-i\gamma})f_{2}^{(t_{2})}.
\end{eqnarray}
Changing the reference point to $t_{3}$, with
$\gamma_{23}=-\gamma$,
\begin{equation}
\psi_{d}=-e^{-2i\gamma}f_{1}^{(t_{3})}-i(1+e^{2i\gamma})f_{2}^{(t_{3})}.
\end{equation}
Going to $e$,
\begin{eqnarray}
\psi_{e}&=&\psi_{d}-i(-i-ie^{2i\gamma})f_{1}^{(t_{3})}\nonumber\\
&=&-(1+e^{2i\gamma}+e^{-2i\gamma})f_{1}^{(t_{3})}-i(1+e^{2i\gamma})f_{2}^{(t_{3})}.
\end{eqnarray}
Going to $f$ and changing the reference point to $t_{4}$, we have
\begin{eqnarray}
\psi_{f}&=&\psi_{e}-i(-1-e^{2i\gamma}-e^{-2i\gamma})f_{2}^{(t_{3})}\nonumber\\
&=&-(1+e^{2i\gamma}+e^{-2i\gamma})e^{i\gamma}f_{1}^{(t_{4})}+ie^{-3i\gamma}f_{2}^{(t_{4})}.
\end{eqnarray}
Going to $g$,
\begin{eqnarray}
\psi_{g}&=&\psi_{f}-i(-1-e^{2i\gamma}-e^{-2i\gamma})e^{i\gamma}f_{2}^{(t_{4})}\nonumber\\
&=&-e^{i\gamma}(1+e^{2i\gamma}+e^{-2i\gamma})f_{1}^{(t_{4})}+
i(e^{3i\gamma}+e^{i\gamma}+e^{-i\gamma}+e^{-3i\gamma})f_{2}^{(t_{4})}.
\end{eqnarray}
Finally we go from $g$ back to $a$ in the counterclockwise
direction at infinity completing the trip around the singularity
point at $r=2M$. Since $f_{1}$ is dominant in this region, the
$f_{1}$ part of $\psi$ does not change. However, there will be an
additional phase contribution, that is,
\begin{equation}
f_{1}^{(t_{4})}=e^{i\tilde{\gamma}_{41}}f_{1}^{(t_{1})},
\end{equation}
with the contour
\begin{equation}
\tilde{\gamma}_{41}+\gamma_{12}+\gamma_{23}+\gamma_{34}=\Gamma\Rightarrow
\tilde{\gamma}_{41}=\Gamma-\gamma,
\end{equation}
where $\Gamma$ is the closed contour integral around $r=2M$ (in
the counterclockwise sense),
\begin{equation}
\Gamma=\oint_{r=2M}d\xi\ Q(\xi)=i4\pi M\omega.\label{gammasch}
\end{equation}
With this phase taken into account, the $f_{1}$ part of $\psi$ at
$\bar{a}$ (back to $a$ after a round trip) is,
\begin{eqnarray}
\psi_{\bar{a}}&=&-e^{i\gamma}(1+e^{2i\gamma}+e^{-2i\gamma})e^{i\Gamma}
e^{-i\gamma}f_{1}^{(t_{1})}+\cdots\nonumber\\
&=&-e^{i\Gamma}(1+e^{2i\gamma}+e^{-2i\gamma})f_{1}^{(t_{1})}+\cdots.
\end{eqnarray}

In the monodromy method of Motl and Neitzke \cite{neitzke}, the
boundary condition at the horizon is translated into the monodromy
requirement of the solution around the singular point at $r=2M$,
\begin{equation}
\psi_{\bar{a}}=e^{-i\Gamma}\psi_{a}.
\end{equation}
Considering only the $f_{1}$ part of the solution $\psi$, and
using the value of $\Gamma$ in Eq.~(\ref{gammasch}), we have
\begin{equation}
-e^{i\Gamma}(1+e^{2i\gamma}+e^{-2i\gamma})=e^{-i\Gamma}\Rightarrow
e^{8\pi M\omega}=-(1+2\ {\rm cos}2\gamma).
\end{equation}

As we can see from Table~\ref{table1}, we have $\gamma=0$ and
$-\pi$ for the scalar and the tensor fields, respectively. In both
cases, we have
\begin{equation}
e^{8\pi M\omega}=-3\Rightarrow\omega=\frac{1}{8\pi M}{\rm ln}3
-\frac{i}{4M}\left(n+\frac{1}{2}\right)
\end{equation}
as $n\rightarrow\infty$. For the Dirac, the Maxwell, and the
Rarita-Schwinger fields, we have $\gamma=-\pi/2$,
\begin{equation}
e^{8\pi M\omega}=1\Rightarrow\omega=-\frac{i}{4M}n\label{SchDirac}
\end{equation}
with zero real part as $n\rightarrow\infty$. This result is
consistent with \cite{jing} and \cite{khriplovich}, but in
contradiction with that in \cite{castello}, where the imaginary
part of the asymptotic quasinormal frequency is found to be
$-in/8M$. In \cite{khriplovich}, the subleading contribution of
lm\ $\omega$ is also calculated. Finally, we note that the spacing
of the imaginary parts of the asymptotic frequencies is $1/4M$ for
all the spin fields.

\section{Reissner-Nordstr\"om black hole}

For the charged Reissner-Nordstr\"om black hole, the form of the
metric is the same as in Eq.~(\ref{metric}), but with
$\Delta=r^{2}-2Mr+q^{2}$ where $q$ is the charge of the black
hole. Since the pole structures in the complex $r$-plane of the
non-extremal and the extremal cases are different, one cannot take
the extremal limit directly from the non-extremal result. We
therefore consider the two cases separately in the following
subsections.

\subsection{Non-extremal case}

\begin{figure}[!]
\includegraphics{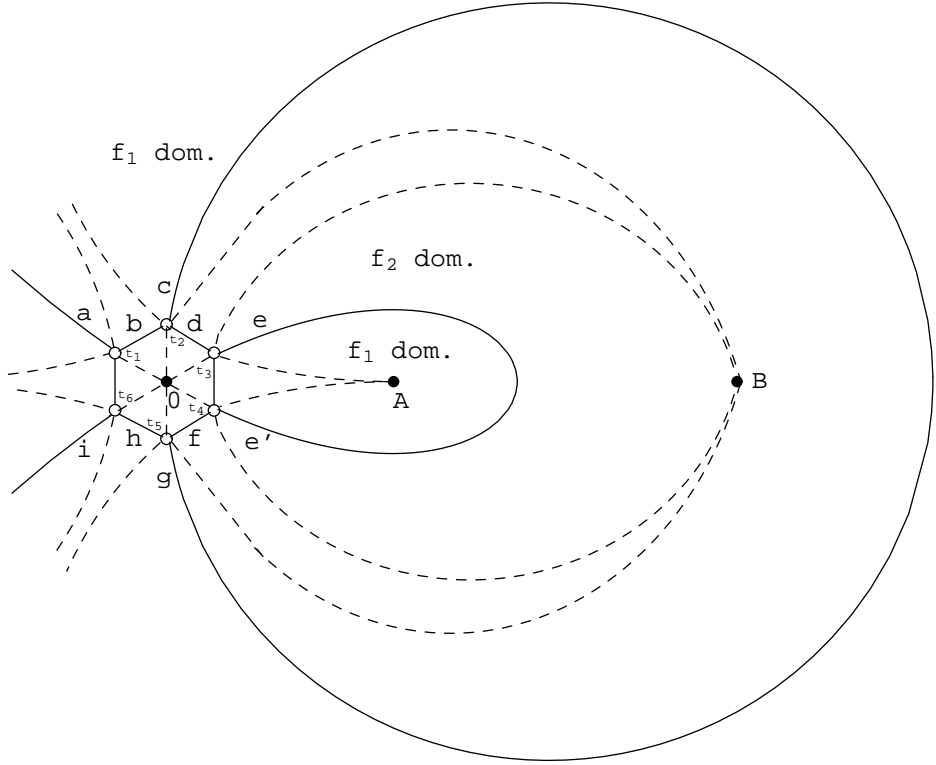}
\caption{\label{RN1}Stokes structure of the non-extreme
Reissner-Nordstr\"om black hole. There are six zeros and three
poles. One pole is at the black hole singularity ($r=0$). The
other two are at the inner horizon $A$ ($r=r_{-}$) and at the
event horizon $B$ ($r=r_{+}$).}
\end{figure}

The wave equation in this case is still given by
Eq.~(\ref{WKBeq}), but with
\begin{equation}
R(r)=\frac{r^{4}}{\Delta^{2}}\left[\omega^{2}-V+\frac{2M}{r^{3}}
-\frac{3(M^{2}+q^{2})}{r^{4}}+\frac{6Mq^{2}}{r^{5}}-\frac{2q^{4}}{r^{6}}\right],
\label{RNR}
\end{equation}
and the WKB solutions are as Eq.~(\ref{WKB}), with
\begin{equation}
Q^{2}(r)=\frac{r^{4}}{\Delta^{2}}\left[\omega^{2}-V-\frac{1}{4r^{2}}+\frac{3M}{r^{3}}
-\frac{(8M^{2}+7q^{2})}{2r^{4}}+\frac{7Mq^{2}}{r^{5}}-\frac{9q^{4}}{4r^{6}}\right].
\label{RNQ}
\end{equation}
There are six zeros and three poles for $Q(r)$. The poles are
located at $r=0$ and $r=r_{\pm}=M\pm\sqrt{M^{2}-q^{2}}$, the event
horizon and the inner horizon, respectively. The corresponding
Stokes structure is indicated in Fig.~\ref{RN1}.

We start with the solution at $a$ again.
\begin{equation}
\psi_{a}=f_{1}^{(t_{1})}.
\end{equation}
We go from $a$ to $b$, to $c$, to $d$, and to $e$, crossing four
anti-Stokes lines. Using the same procedure as in the last
section, we obtain
\begin{equation}
\psi_{e}=-(1+e^{2i\gamma}+e^{-2i\gamma})f_{1}^{(t_{3})}
-i(1+e^{2i\gamma})f_{2}^{(t_{3})},
\end{equation}
where
$\gamma_{12}=-\gamma_{23}=\gamma_{34}=-\gamma_{45}=\gamma_{56}=\gamma$
with
\begin{eqnarray}
\gamma&=&\int_{t_{1}}^{t_{2}}d\xi\ Q(\xi)
=\int_{t_{1}}^{t_{2}}d\xi\ \left(\frac{\xi^{2}}{q^{2}}\right)
\left[\omega^{2}-\frac{(9+4\alpha)q^{4}}{4\xi^{6}}\right]^{1/2}\nonumber\\
&=&\frac{\pi}{2}\sqrt{1+\frac{4\alpha}{9}}.\label{RNgamma}
\end{eqnarray}
We assume here that the asymptotic behavior of the effective
potential is
\begin{equation}
V|_{r\rightarrow 0}\sim\frac{q^{4}}{r^{6}}\alpha.\label{RNV}
\end{equation}
The value of $\alpha$ and $\gamma$ for various spin fields are
tabulated in Table~\ref{table2}.

\begin{table}
\caption{\label{table2}Some parameters characterizing the
asymptotic behaviors of various spin fields for the
Reissner-Nordstr\"om black hole. $\alpha$ represents the behavior
of the effective potential near the black hole singularity as in
Eq.~(\ref{RNV}). $\gamma$ is the value of the integral in
Eq.~(\ref{RNgamma}). }
\begin{ruledtabular}
\begin{tabular}{ccccc}
spin&$\alpha$&$\gamma$&$-2(1+{\rm cos}2\gamma)$&$1+2\ \!{\rm
cos}2\gamma+2\ \!{\rm cos}4\gamma$\\ \hline 0 & $-2$ & $\pi$/6 &
$-3$ & 1
\\ 1/2 & 0 & $\pi$/2 & 0 & 1
\\ 1 & 4 & $5\pi$/6 & $-3$ & 1
\\ 3/2 & $-2$ & $\pi$/6 & $-3$ & 1 \\ 2 & 4 & 5$\pi$/6 & $-3$ & 1 \\
\end{tabular}
\end{ruledtabular}
\end{table}

To circle the singular point at $r=r_{+}$, we go along the
anti-Stokes line from $e$ to $e'$. The behavior of $\psi$ does not
change, but there will be additional phases. The phase
$\tilde{\gamma}_{34}$ satisfies
\begin{equation}
\tilde{\gamma}_{34}+\gamma_{43}=-\Gamma_{-}\Rightarrow
\tilde{\gamma}_{34}=-\Gamma_{-}+\gamma,
\end{equation}
where
\begin{equation}
\Gamma_{-}=\oint_{r=r_{-}}d\xi\ Q(\xi)=-i\pi\omega
M\frac{(1-\kappa)^{2}}{\kappa},
\end{equation}
with $\kappa=\sqrt{1-q^{2}/M^{2}}$. Since we are considering the
non-extremal case, we have $0<\kappa\leq 1$. Hence, with the
contribution of the phase $\tilde{\gamma}_{34}$,
\begin{eqnarray}
\psi_{e'}&=&-(1+e^{2i\gamma}+e^{-2i\gamma})e^{i\tilde{\gamma}_{34}}f_{1}^{(t_{4})}
-i(1+e^{2i\gamma})e^{-i\tilde{\gamma}_{34}}f_{2}^{(t_{4})}\nonumber\\
&=&-e^{-i\Gamma_{-}}e^{i\gamma}(1+e^{2i\gamma}+e^{-2i\gamma})f_{1}^{(t_{4})}
-ie^{i\Gamma_{-}}(e^{i\gamma}+e^{-i\gamma})f_{2}^{(t_{4})}.
\end{eqnarray}

Going from $e'$ to $f$, to $g$, to $h$, and then to $i$, we have
\begin{eqnarray}
\psi_{i}&=&\left[-e^{i\Gamma_{-}}e^{i\gamma}(2+e^{2i\gamma}+e^{-2i\gamma})
-e^{-i\Gamma_{-}}e^{i\gamma}(1+e^{2i\gamma}+e^{-2i\gamma})\right]f_{1}^{(t_{6})}\nonumber\\
&&\ \
+\left[ie^{i\Gamma_{-}}(e^{3i\gamma}+2e^{i\gamma}+2e^{-i\gamma}+e^{-3i\gamma})
+ie^{-i\Gamma_{-}}(e^{i\gamma}+e^{-i\gamma})(1+e^{2i\gamma}+e^{-2i\gamma})\right]f_{2}^{(t_{6})}.
\nonumber\\
\end{eqnarray}
To circle the singular point at $r=r_{+}$, we go back to $a$ at
infinity in the counterclockwise direction. The phase contribution
satisfies
\begin{equation}
\tilde{\gamma}_{61}+\gamma_{16}=\Gamma_{+}+\Gamma_{-}\Rightarrow
\tilde{\gamma}_{61}=\Gamma_{+}+\Gamma_{-}-\gamma,
\end{equation}
where
\begin{equation}
\Gamma_{+}=\oint_{r=r_{+}}d\xi\ Q(\xi)=i\pi\omega
M\frac{(1+\kappa)^{2}}{\kappa}.
\end{equation}
Considering only the $f_{1}$ part of the solution,
\begin{eqnarray}
\psi_{\bar{a}}&=&\left[-e^{i\Gamma_{-}}e^{i\gamma}(2+e^{2i\gamma}+e^{-2i\gamma})
-e^{-i\Gamma_{-}}e^{i\gamma}(1+e^{2i\gamma}+e^{-2i\gamma})\right]e^{i\Gamma_{+}}
e^{i\Gamma_{-}}e^{-i\gamma}f_{1}^{(t_{1})}+\cdots\nonumber\\
&=&-e^{i\Gamma_{+}}\left[e^{2i\Gamma_{-}}(2+e^{2i\gamma}+e^{-2i\gamma})
+(1+e^{2i\gamma}+e^{-2i\gamma})\right]f_{1}^{(t_{1})}+\cdots
\end{eqnarray}
The monodromy requirement corresponding to the boundary condition
at the event horizon $r=r_{+}$ is
\begin{equation}
\psi_{\bar{a}}=e^{-i\Gamma_{+}}\psi_{a}.
\end{equation}
Therefore, we have
\begin{eqnarray}
&&-e^{i\Gamma_{+}}\left[e^{2i\Gamma_{-}}(2+e^{2i\gamma}+e^{-2i\gamma})
+(1+e^{2i\gamma}+e^{-2i\gamma})\right]=e^{-i\Gamma_{+}}\nonumber\\
&\Rightarrow&e^{-2i\Gamma_{+}}=1-2(1+{\rm
cos}2\gamma)(1+e^{2i\Gamma_{-}}).
\end{eqnarray}

From Table~\ref{table2}, we have for the spin 0, the spin 1, and
the spin 2 fields, $-2(1+{\rm cos}2\gamma)=-3$, so
\begin{equation}
e^{-2i\Gamma_{+}}=-2-3e^{2i\Gamma_{-}}\Rightarrow e^{2\pi\omega
M(1+\kappa)^{2}/\kappa}=-2-3e^{2\pi\omega M(1-\kappa)^{2}/\kappa}.
\label{RNQNM}
\end{equation}
Hence, these three fields have the same asymptotic frequency
although it cannot be written in a close form as in the
Schwarzschild case. Our result agrees with that in Refs.
\cite{neitzke,natario,howls}. For the spin 3/2 Rarita-Schwinger
field, the same result is obtained because the corresponding
$\alpha$ and $\gamma$ are identical to that of the integral spin
fields. The only field that has a different asymptotic frequency
in this case is the Dirac field. For the Dirac field, we have
$2(1+{\rm cos}2\gamma)=0$ as listed in Table~\ref{table2} and
\begin{equation}
e^{-2i\Gamma_{+}}=1\Rightarrow\omega=-in\frac{\kappa}{M(1+\kappa)^{2}}.
\label{RNDirac}
\end{equation}
The real part of the Dirac asymptotic frequency is zero, and the
spacing of the imaginary part varies with $q$.

\subsection{Extremal case}

Suppose we naively take the extremal limit $q\rightarrow M$ or
$\kappa\rightarrow 0$ of the result in Eq.~(\ref{RNQNM}) in the
last subsection, we obtain
\begin{equation}
e^{8\pi\omega M}=-3.
\end{equation}
for spin 0, 1, 3/2, and 2 fields. This coincides with the result
for the scalar and the tensor fields for the Schwarzschild black
hole. However, for $s=1$ and $3/2$, it differs from the
Schwarzschild result (Eq.~(\ref{SchDirac})). Moreover, for the
Dirac case in Eq.~(\ref{RNDirac}), the limit $\kappa\rightarrow 0$
is in fact inconsistent. As already pointed out in \cite{howls},
one cannot take this limit directly to obtain the extremal result.
There are only two poles in the extremal case, while in the
non-extremal case there are three. The structure of the complex
plane is different and a separate analysis has to be carried out.
This is what we shall do next.

\begin{figure}[!]
\includegraphics{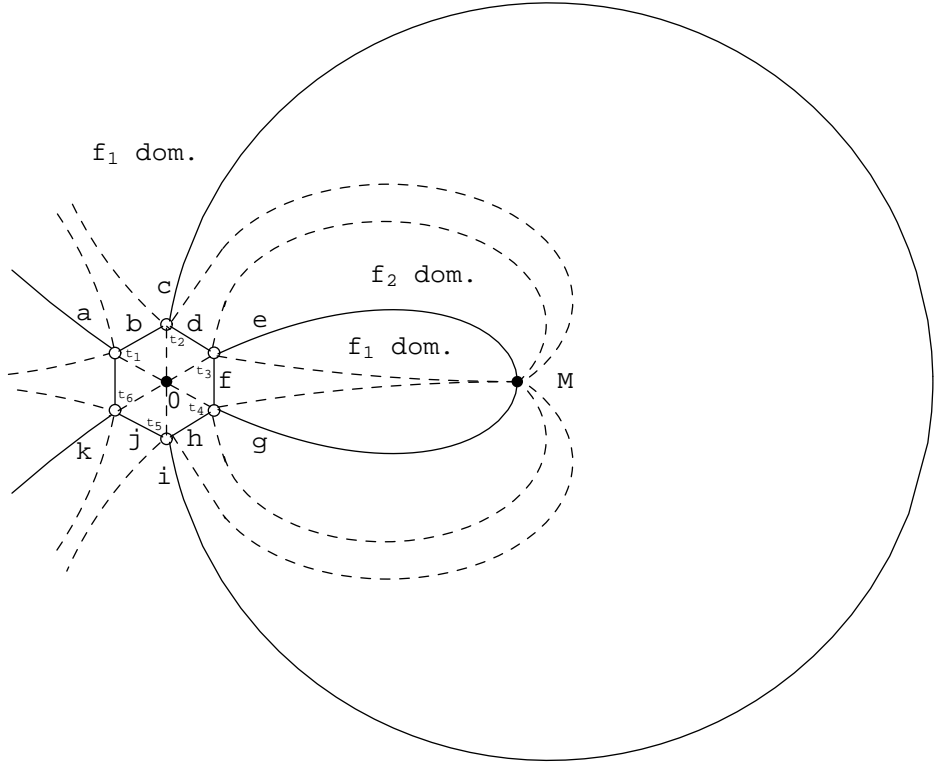}
\caption{\label{RN2}Stokes structure of the extreme
Reissner-Nordstr\"om black hole. There are only two poles here.
One is at the black hole singularity ($r=0$) and the other one is
at the event horizon ($r=M$).}
\end{figure}

The functions $R(r)$ and $Q(r)$ in this case can be obtained by
taking the extremal limit $q\rightarrow M$ in Eqs.~(\ref{RNR}) and
(\ref{RNQ}).
\begin{equation}
R(r)=\frac{r^{4}}{(r-M)^{4}}\left[\omega^{2}-V+\frac{2M}{r^{3}}
-\frac{6M^{2}}{r^{4}}+\frac{6M^{3}}{r^{5}}-\frac{2M^{4}}{r^{6}}\right],
\end{equation}
and
\begin{equation}
Q^{2}(r)=\frac{r^{4}}{(r-M)^{4}}\left[\omega^{2}-V-\frac{1}{4r^{2}}+\frac{3M}{r^{3}}
-\frac{15M^{2}}{2r^{4}}+\frac{7M^{3}}{r^{5}}-\frac{9M^{4}}{4r^{6}}\right].
\end{equation}
There are again six zeros but only two poles for $Q(r)$. At the
horizon, $r=M$, we have a double pole, which locates right on one
of the Stokes lines. The corresponding Stokes structure is shown
in Fig.~\ref{RN2}.

Going through the same analysis as before, we start at $a$,
\begin{equation}
\psi_{a}=f_{1}^{(t_{1})}.
\end{equation}
We then go to $b$, $c$, $\dots$, $j$, and $k$ in the
counterclockwise direction, and we have
\begin{equation}
\psi_{k}=(e^{5i\gamma}+e^{3i\gamma}+e^{i\gamma}+e^{-i\gamma}+e^{-3i\gamma})f_{1}^{(t_{6})}
-i(e^{5i\gamma}+e^{3i\gamma}+e^{i\gamma}+e^{-i\gamma}+e^{-3i\gamma}+e^{-5i\gamma})f_{2}^{(t_{6})},
\end{equation}
where
\begin{equation}
\gamma=\frac{\pi}{2}\sqrt{1+\frac{4\alpha}{9}},
\end{equation}
as given in Eq.~(\ref{RNgamma}) in the non-extremal case with the
asymptotic behavior of the effective potential
\begin{equation}
\left.V\right|_{r\rightarrow 0}\sim\frac{M^{4}}{r^{6}}\alpha,
\end{equation}
which is also the same as in the non-extremal case.

Circling the singular point at $r=M$ back to $a$ at infinity as
before, we have
\begin{equation}
\tilde{\gamma}_{61}+\gamma_{16}=\Gamma\Rightarrow
\tilde{\gamma}_{61}=\Gamma-\gamma,
\end{equation}
where
\begin{equation}
\Gamma=\oint_{r=M}d\xi\ Q(\xi)=i4\pi M\omega.
\end{equation}
Therefore,
\begin{eqnarray}
\psi_{\bar{a}}&=&(e^{5i\gamma}+e^{3i\gamma}+e^{i\gamma}+e^{-i\gamma}+e^{-3i\gamma})
e^{i\Gamma}e^{-i\gamma}f_{1}^{(t_{1})}+\cdots\nonumber\\
&=&e^{i\Gamma}(e^{4i\gamma}+e^{2i\gamma}+1+e^{-2i\gamma}+e^{-4i\gamma})f_{1}^{(t_{1})}+\cdots.
\end{eqnarray}
The monodromy condition is again
\begin{eqnarray}
&&e^{i\Gamma}(e^{4i\gamma}+e^{2i\gamma}+1+e^{-2i\gamma}+e^{-4i\gamma})=e^{-i\Gamma}\nonumber\\
&\Rightarrow&e^{-2i\Gamma}=1+2\ \!{\rm cos}2\gamma+2\ \!{\rm
cos}4\gamma.
\end{eqnarray}

For all the spin fields, as shown in Table~\ref{table2},
\begin{equation}
1+2\ \!{\rm cos}2\gamma+2\ \!{\rm cos}4\gamma=1.
\end{equation}
Hence, the asymptotic quasinormal frequency in the extremal case
for all the spin fields is
\begin{equation}
e^{8\pi M\omega}=1\Rightarrow\omega=-in\left(\frac{1}{4M}\right).
\end{equation}
It is thus curious to see that all the different spin fields have
the same asymptotic frequency for the extreme black hole. This
result is consistent with that in \cite{natario} where the
integral spin cases are considered.

\section{Conclusions and Discussions}

\begin{table}
\caption{\label{table3}Quasinormal frequencies of various spin
fields for the Schwarzschild and the Reissner-Nordstr\"om (RN)
black holes (BH). Eq.~(\ref{RNQNM}) is the equation that
determines the frequencies in some Reissner-Nordstr\"om cases
where the frequencies cannot be written in a close form.
$\kappa=\sqrt{1-q^{2}/M^{2}}$.}
\begin{ruledtabular}
\begin{tabular}{cccc}
spin&Schwarzschild BH&RN BH&Extreme RN BH\\ \hline 0 &
$\frac{1}{8\pi M}{\rm ln}3
-\frac{i}{4M}\left(n+\frac{1}{2}\right)$  & Eq.~(\ref{RNQNM}) &
$-\frac{i}{4M}n$
\\ 1/2 & $-\frac{i}{4M}n$ &  $-\frac{i\kappa}{(1+\kappa)^{2}M}n$ & $-\frac{i}{4M}n$
\\ 1 &
$-\frac{i}{4M}n$ & Eq.~(\ref{RNQNM}) & $-\frac{i}{4M}n$
\\ 3/2 & $-\frac{i}{4M}n$ & Eq.~(\ref{RNQNM}) & $-\frac{i}{4M}n$  \\ 2 &
$\frac{1}{8\pi M}{\rm ln}3
-\frac{i}{4M}\left(n+\frac{1}{2}\right)$ & Eq.~(\ref{RNQNM}) &
$-\frac{i}{4M}n$ \\
\end{tabular}
\end{ruledtabular}
\end{table}

We have evaluated the asymptotic quasinormal frequencies for
various spin fields in Schwarzschild and Reissner-Norstr\"om black
holes, using a combination of the monodromy method of Motl and
Neitzke \cite{neitzke} and the WKB formalism of Andersson and
Howls \cite{howls}. These frequencies are tabulated in
Table~\ref{table3}. In the Schwarzschild case, the real part of
the asymptotic frequency for the spin 0 and the spin 2 fields is
ln3. This value has inspired a lot of interesting in its relation
to the black hole area and entropy spectra \cite{hod1}. However,
the real part of the frequency for the spin 1/2, the spin 1, and
the spin 3/2 fields is zero. This result casts doubts on the
universality of the value of ln3, even for single-horizon black
holes. On the other hand, the imaginary parts of the frequencies
all have spacings $1/4M$ or $2\pi T_{S}$, where $T_{S}$ is the
Hawking temperature of the Schwarzschild black hole. This value is
thus universal for all the spin fields \cite{padmanabhan,medved}.

Our result for the imaginary part of the Dirac quasinormal
frequency agrees with \cite{jing} and \cite{khriplovich}, but in
contradiction with that of \cite{castello}. In \cite{castello},
the imaginary part of the frequency is calculated in two different
ways, one analytical and the other numerical. For the analytical
calculation, the authors there follow the method of
\cite{padmanabhan} and \cite{medved} in which the imaginary part
is derived by identifying the locations of the poles of the
scattering amplitude in the Born approximation. This calculation
is criticized in \cite{jing} where it is shown that the method of
\cite{padmanabhan} and \cite{medved} for the integral spin fields
cannot be extended directly to the Dirac case. Hence, the validity
of the analytical calculation is in question. As for the numerical
analysis, the authors use the continued fraction method of Leaver
\cite{leaver} which converges much slower than the modified method
of Nollert \cite{nollert}. It seems that the method of Nollert
cannot be applied to the effective potential of the Dirac field.
It is therefore possible that the correct answer has not been
reached numerically there. In any case, the conclusion on the
spacing of the imaginary part of the frequencies in
\cite{castello} is not at all reliable.

For the non-extreme Reissner-Nordstr\"om black holes, we find that
the asymptotic frequency of the spin 3/2 Rarita-Schwinger field is
the same as that of the spin 0, the spin 1, and the spin 2 fields
which was first evaluated in \cite{motl,neitzke}. Since this
frequency involves both the mass $M$ and the charge $q$ of the
black hole, it cannot be expressed in a close form as that of the
Schwarzschild case \cite{howls}. Recently, Hod \cite{hod2} had
taken up this problem again. He obtained a value of ln2 for the
real part of the frequency even in the Reissner-Nordstr\"om case
by considering the quasinormal modes of a charged scalar field. It
would be interesting to see if a universal value can be obtained
for other spin fields by extending our calculation to charged
field cases.

The spin 1/2 Dirac field is special in the Reissner-Nordstr\"om
case. Its asymptotic quasinormal frequency is different from the
other spin fields. The real part is zero, as in the Schwarzschild
case. The imaginary part has a spacing of $\kappa/
M(1+\kappa)^{2}$ or $2\pi T_{RN}$, where $T_{RN}$ is the Hawking
temperature of the Reissner-Nordstr\"om black hole. In terms of
the Hawking temperature, this spacing has the same form as that
for the Schwarzschild black hole.

In the extremal case, it is curious to see that all the spin
fields have the same asymptotic frequency. Since the extreme
Reissner-Nordstr\"om black hole is the simplest example of an
supersymmetric black hole, one would expect the spin 1, the spin
3/2, and the spin 2 fields to have the same asymptotic frequency
\cite{onozawa}. However, it is quite unexpected for the scalar and
the Dirac fields to have the same value. The real part of this
frequency is zero, while the imaginary part has again a spacing of
$1/4M$. This spacing cannot be expressed in terms of the Hawking
temperature which is zero in the extremal case, but it is the same
as the spacing in the Schwarzschild case. In this respect, it is
the spacing of the imaginary part of the asymptotic frequencies
that has a universal value of $1/4M$ for all the single-horizon
cases of the Schwarzschild and the extreme Reissner-Nordstr\"om
black holes.

We can see from above that the universality of ln3 is indeed in
question even for the single-horizon black holes. In addition, the
relevancy of the asymptotic quasinormal frequency to the
microstate description of the black hole entropy is not clear
\cite{daghigh} because this frequency depends crucially on the
behavior of the effective potential near the black hole
singularity as well as that near the event horizon. In spite of
this, we still think that the black hole quasinormal spectrum
should be important in the understanding of the quantum properties
of the black hole \cite{york1,york2}. The quasinormal modes
represent the characteristic oscillations of the black hole. If
they are quantized in an appropriate way, which would involve the
problem of how to quantize an open system, we should be able to
obtain more information on the entropy of the quantum black hole
\cite{kim}.

Finally, it would be desirable to extend our consideration to the
case of the Kerr black hole in order to have a further
understanding of the universality question of the asymptotic
quasinormal frequencies. Up to now, the evaluations of the Kerr
asymptotic frequencies are mostly numerical (see, for example,
\cite{hod3,berti1,berti2,musiri,setare,hod4}). This is because one
has to deal with the asymptotic behaviors of the radial equation
as well as the angular equation, which involves the spheroidal
harmonics. Recently there are a number of studies on the
asymptotic behaviors of the spheroidal harmonics
\cite{casals,berti3}. Hopefully one would soon be able to carry
out a more complete study of the asymptotic quasinormal
frequencies in the Kerr black hole case.

\begin{acknowledgments}
The author would like to thank the hospitality of Kin-Wang Ng and
the Theory Group of the Institute of Physics at the Academia
Sinica, Taiwan, Republic of China, where this work was initiated.
This research is supported by the National Science Council of the
Republic of China under contract number NSC 94-2112-M-032-009.
\end{acknowledgments}

\end{document}